\documentstyle[11pt]{article}

\title{\bf The bootstrap condition for many reggeized gluons and the
photon structure function at low $x$ and $N_{c}\rightarrow\infty$}
\author{M.Braun \\ Department of high-energy
physics, \\ University of S. Petersburg, 198904 S. Petersburg
, Russia and\\
Department of Particle Physics, University of Santiago de Compostela,\\
15706 Santiago de Compostela, Spain
}
\def\beq{\begin{equation}}
\def\eeq{\end{equation}}
\def\noi{\noindent}

\begin{document}
\maketitle
\medskip
\noi{\bf Abstract.}

The bootstrap condition is generalized to $n$ reggeized gluons. As a
result it is demonstrated that the intercept generated by $n$ reggeized
gluons cannot be lower than the one for $n=2$. Arguments are presented
that in the limit $N_{c}\rightarrow\infty$ the bootstrap condition
reduces the $n$ gluon chain with interacting neighbours to a single
BFKL pomeron. In this limit the leading contribution from $n$ gluons
corresponds to $n/2$ non-interacting BFKL pomerons (the $n/2$ pomeron
cut). The sum over $n$ leads to a unitary $\gamma^{\ast}\gamma$
amplitude of the eikonal form.
\vspace{3 cm}

{\Large\bf SPbU-IP-1995/3}
\newpage
\section{Introduction.}

Recently much attention has been given to the system of many reggeized
gluons with a pair interaction between them in the framework of the
equation proposed by J.Bartels, J. Kwiecinski and M.Praszalowicz [ 1,2
]. Mostly the case of infinite number of colours, $N_{c}\rightarrow\infty$,
was investigated, when the colour structure of the equation drastically
simplifies. The resulting chain Hamiltonian, with only the neighbours
interacting, has been extensively studied in [ 3-6 ]. The conformal
symmetry seems to reduce the problem to a completely integrable one but
it still remains very complicated.

A close study of the old results on the two-gluon case and the results
by J.Bartels on the three-and four-gluon cases [ 7 ] suggests, however,
that the physical solutions depend heavily on the form of the
inhomogeneous term in the equation, that is, the gluon coupling to the
target (projectile). It may occur that although the complete spectrum of
the $n$-gluon Hamiltonian is complicated and hardly calculable, the
physical solutions corresponding to a given coupling to the external
source are simple and easy to find. It is this aspect of the $n$-gluon
problem, which is studied in the present paper. We do not touch here the
problem of the introduction of the running coupling constant into the
formalism. The coupling constant is assumed to be fixed and
small throughout the paper.

To have a well-defined external gluon source, we restrict ourselves to
the interaction of photons, real or virtual. A photon splits into a
$q\bar{q}$ pair which emits gluons in a well-defined and calculable
manner. In the lowest order in the coupling constant it corresponds to
the coupling of $n$ gluons to a single $q\bar{q}$ loop. For this
particular target (projectile) we argue that in the limit
$N_{c}\rightarrow\infty$ the physical solutions in the $n$-gluon system
are quite simple. They separate into two classes depending on the colour
structure. One class of solutions corresponds to the gluons pairing into
colour singlets. Then the interaction between different pairs vanishes
in the limit $N_{c}\rightarrow\infty$ and the solution reduces to the
product of $n/2$ independent BFKL pomerons [ 8 ], i.e. to the standard
$n/2$ pomeron cut. The other solution corresponds to the gluons being
in the vector (adjoint) colour representation with their neighbours.
It is for
this colour configuration that the Hamiltonian aquires the chain form
studied in [ 3-6 ]. Our result is that in this case, due to the
so-called bootstrap condition discovered for two gluons in [ 9 ]
and generalized  for $n$ gluons here, all gluons coalesce into a single
pair, that is, the solution for $n$ gluons reduces to a single BFKL
pomeron. To clarify the meaning of this result, we stress that by no
means do we assert that all eigenstates of the corresponding Hamiltonian
have this structure. Our result is that only a particular class of
solutions, which are precisely the ones of physical interest, reduce to
a single pomeron.

We have also to stress that our results in this respect rest heavily on
those by J.Bartels, who studied the system  of four reggeized gluons
with $N_{c}=3$ in [ 7 ]. Our investigation was, in fact, inspired by
this paper. We also mention the results by G.Korchemsky [ 6 ]
who using a sophisticated one-dimensional lattice technique conjectured
that the minimal energy of the $n$ gluon lattice coincides with the one
for $n=2$.

\section{The bootstrap condition for arbitrary number of gluons}

We start by recalling the bootstrap condition for two reggeized gluons
as obtained in [ 9 ]. The two-gluon equation can be written as an
inhomogeneous Schroedinger equation
\beq
(H-E)\psi =F
\eeq
with the Hamiltonian for the colour group $SU(N_{c})$
\beq
H=(N_{c}/2)(t(q_{1})+t(q_{2}))+(T_{1}T_{2})V
\eeq
Here $(N_{c}/2)t(q)$ is the kinetic energy of the gluon given by its
Regge trajectory
\beq
t(q)=g^{2}\int(d^{2}q_{1}/(2\pi)^{3})\frac{q^{2}}{q_{1}^{2}q_{2}^{2}},\
\ q_{1}+q_{2}=q
\eeq
Infrared regularization is understood here and in the following where
necessary (either dimensional or by a nonzero gluon mass). The
interaction term, apart from the product of the gluon colour vectors
$T_{i}^{a},\ i=1,2,\ a=1,...N_{c}$, involves the BFKL kernel [ 8 ]
\beq
(1/g^{2})V(q_{1},q_{2},q'_{1},q'_{2})=
(\frac{q_{1}^{2}}{{q'_{1}}^{2}}+\frac{q_{2}^{2}}{{q'_{2}}^{2}})
\frac{1}{(q_{1}-q'_{1})^{2}}-\frac{q^{2}}{{q'_{1}}^{2}{q'_{2}}^{2}}
\eeq
with $q_{1}+q_{2}=q'_{1}+q'_{2}$. Comparing (3) with (4) one finds the
bootstrap condition
\beq
\int(d^{2}q_{1}/(2\pi)^{3})V(q_{1},q_{2},q'_{1},q'_{2})=
t(q_{1})+t(q_{2})-t(q_{1}+q_{2})
\eeq
It is customary to discuss a homogeneous equation (1) (with $F=0$) to
seek for the ground state energy, which determines the rightmost
singularity in the complex angular momentum $j$ related to the energy by
\beq j=1-E \eeq
For our purpose it is, however, essential to conserve the inhomogeneous
term $F(q_{1},q_{2})$, which represents the two-gluon-particle vertex.

Consider the vector colour channel with $T_{1}T_{2}=-N_{c}/2$ and, most
important, assume that the inhomogeneous term depends only on the total
momentum of the two gluons: $F=F(q_{1}+q_{2})$. Then using the bootstrap
relation (5) one easily finds the solution to Eq. (1):
\beq
\psi(q_{1},q_{2})=\psi(q_{1}+q_{2})=F(q_{1}+q_{2})/(t(q_{1}+q_{2})-E)
\eeq
It means that the two gluons 1 and 2 have fused into a single one with
the momentum $q_{1}+q_{2}$.

We have to stress two points important for the following. First, the
solution (7) is the correct and unique solution for the given form of
the
inhomogeneous term $F=F(q_{1}+q_{2})$. It does not involve all possible
eigenstates of the Hamiltonian, which may be quite complicated and which
might appear should the inhomogeneous term depend on $q_{1}$ and $q_{2}$
separately. Still if we know that the external coupling has a specific
dependence $F(q_{1}+q_{2})$ the found solution is the desired one and
we need not bother about other eigenstates of the Hamiltonian. Second,
the solution refers to both signatures, positive and negative. The
solution with the negative signature physically represents the reggeized
gluon itself. It is this solution which is meant by the bootstrap.
However the same singularity in $j$ appears also in the positive
signature where it does not correspond to any particle at $j=1$. We thus
observe degeneration in signature.

We pass now to the main topic of this section:  a generalization of this
result to the $n$-gluon case. Fo $n$ gluons Eq. (1) holds with the
Hamiltonian which is a sum of kinetic energies and pair interactions:
\beq
H=(N_{c}/2)\sum_{i=1}^{n}t(q_{i})+\sum_{i<k}^{n}(T_{i}T_{k})V_{ik}
\eeq
Here $V_{ik}$ is the interaction of gluons $i$ and $k$ with the kernel
$V(q_{i},q_{k},q'_{i},q'_{k})$. We do not impose any restrictions on the
total colour $T=\sum_{i=1}^{n}T_{i}$. (However only for $T=0$ the
Hamiltonian is infrared stable).

Assume now that $T_{1}T_{2}=-N_{c}/2$, i.e. the gluons 1 and 2 form a
colour vector. Then we claim that for a certain specific choice of the
inhomogeneous term $F$  Eq. (1) for $n$ gluons reduces to that for $n-1$
gluons, the gluons 1 and 2 fused into a single gluon which carries their
total momentum and colour. In other words, as in the two-gluon case, a
pair
of gluons in the adjoint representation is equivalent to a single gluon.

Of course, a specific form of the inhomogeneous term is a decisive
instrument for bootstrapping the two gluons 1 and 2. For $n$ gluons we
choose
\[
F_{n}(q_{1},q_{2},...,q_{n})=\sum_{i=3}^{n}\int (d^{2}q'_{i}/(2\pi)^{3})
\]\beq
\hat{W}(q_{1},q_{2},q_{i};q'_{1}q'_{i})\psi
(q'_{1},q_{3},...q'_{i},...q_{n})+F_{n-1}(q_{1}+q_{2},q_{3},...q_{n})
\eeq
with $q'_{1}+q'_{i}=q_{1}+q_{2}+q_{i}$. In Eq. (9) $\psi_{n-1}$ is the
solution of the Schroedinger equation (1) with the inhomogeneous term
$F_{n-1}$. Gluon 1 in it substitutes the fused initial gluons 1 and 2.
The kernel $\hat{W}$ is also an operator acting on colour indeces of
$\psi_{n-1}$. It is convenient to retain a pair of colour indeces for
gluon 1 in $\psi_{n-1}$ inherited from the initial gluons 1 and 2 by
means of a projector onto the adjoint representation in the colour
$T_{1}+T_{2}$. This allows to consider $\hat{W}$ as an operator acting
in the colour space of $n$ gluons. Then it has the form
\beq
\hat{W}(q_{1},q_{2},q_{i};q'_{1},q'_{i})=
(T_{1}T_{i})W(q_{1},q_{2},q_{i};q'_{1},q'_{i})+
(T_{2}T_{i})W(q_{2},q_{1},q_{i};q'_{1},q'_{i})
\eeq
the momentum space kernel $W$ being a difference between two BFKL
kernels
\beq
(1/g)W(q_{1},q_{2},q_{i};q'_{1},q'_{i})=
V(q_{1}+q_{2},q_{i},q'_{1},q'_{i})-V(q_{1},q_{i},q'_{1}-q_{2},q'_{i})
\eeq

Let us demonstrate that with the inhomogeneous term $F_{n}$ given by Eq.
(9) and $T_{1}T_{2}=-N_{c}/2$ the Schroedinger equation (1) is solved by
\beq
\psi_{n}(q_{1},q_{2},q_{3},....q_{n})
=g\psi_{n-1}(q_{1}+q_{2},q_{3},....q_{n})
\eeq
Indeed putting (12) into the equation we find that the interaction term
$(T_{1}T_{2})V_{12}$, according to (5), substitutes the sum of kinetic
terms $t(q_{1})+t(q_{2})$ for the gluons 1 and 2 by $t(q_{1}+q_{2})$
which
is precisely the kinetic term for the function $\psi_{n-1} (q_{1}+q_{2},
q_{3},...)$. The interaction of gluon 1 with the $i$th one, $i\geq 3$,
takes the form
\[
(T_{1}T_{i})\int (d^{2}q'_{i}/(2\pi)^{3})V(q_{1},q_{i},q'_{1},q'_{i})
g\psi_{n-1}(q'_{1}+q_{2},q_{3},...q'_{i},...q_{n})=\]
\beq (T_{1}T_{i})\int
(d^{2}q'_{i}/(2\pi)^{3})V(q_{1},q_{i},q'_{1}-q_{2},q'_{i})
g\psi_{n-1}(q'_{1},q_{3},...q'_{i},...q_{n})
\eeq
The momentum is conserved during the interaction, so that on
the lefthand side $q_{1}+q_{i}=q'_{1}+q'_{i}$ and on the righthand side
$q_{1}+q_{2}+q_{i}=q'_{1}+q'_{i}$.  The term (13) is cancelled by the
identical contribution with the opposite sign coming from the
inhomogeneous term $F_{n}$ (the second term in (11) in the part of
$\hat{W}$ proportional to $T_{1}T_{i}$). Instead of it from the
inhomogeneous term comes the contribution (the first term in (11))
\beq
(T_{1}T_{i})
\int (d^{2}q'_{i}/(2\pi)^{3})V(q_{1}+q_{2},q_{i},q'_{1},q'_{i})
g\psi_{n-1}(q'_{1},q_{3},...q'_{i},...q_{n})
\eeq
In the same manner the interaction of gluon 2 with the $i$th one
\[
(T_{2}T_{i})\int (d^{2}q'_{i}/(2\pi)^{3})V(q_{2},q_{i},q'_{2},q'_{i})
g\psi_{n-1}(q_{1}+q'_{2},q_{3},...q'_{i},...q_{n})=\]
\beq (T_{2}T_{i})\int
(d^{2}q'_{i}/(2\pi)^{3})V(q_{2},q_{i},q'_{2}-q_{1},q'_{i})
g\psi_{n-1}(q'_{2},q_{3},...q'_{i},...q_{n})
\eeq
is transformed by the term in $F_{n}$ proportional to $T_{2}T_{i}$ into
\beq
(T_{2}T_{i})
\int (d^{2}q'_{i}/(2\pi)^{3})V(q_{1}+q_{2},q_{i},q'_{1},q'_{i})
g\psi_{n-1}(q'_{1},q_{3},...q'_{i},...q_{n})
\eeq
The two terms (14) and (16) sum into
\beq
(T_{1}+T_{2},T_{i})
\int (d^{2}q'_{i}/(2\pi)^{3})V(q_{1}+q_{2},q_{i},q'_{1},q'_{i})
g\psi_{n-1}(q'_{1},q_{3},...q'_{i},...q_{n})
\eeq
which is precisely the correct form for the interaction of the gluon
with the total colour $T_{1}+T_{2}$ and momentum $q_{1}+q_{2}$
in which the initial gluons 1 and 2 have fused. Other kinetic energies
$t(q_{i})$ and interactions $(T_{i}T_{k})V_{ik}$  with $i,k\geq 3$ are
not influenced by the bootstrap of gluons 1 and 2. As a result we obtain
the Schroedinger equation (1) for the function $\psi_{n-1}$ describing
$n-1$ gluons with the inhomogeneous term $F_{n-1}$.

The obtained result immediately allows to conclude that the leading
singularity in the complex momentum $j$ for the system of $n$ gluons
(the ground state energy) cannot in any case be lower (higher) than for
the system of $n-1$ gluons. In fact, for any solution for $n-1$ gluons
we can construct a solution for $n$ gluons using $F_{n}$ given by (8)
as the inhomogeneous term. Note that $F_{n}$ depends on energy $E$ via
the solution $\psi_{n-1}$ and thus posseses all branch points in $E$
which mark thresholds in the spectrum for $n-1$ gluons. The fact that
the final solution $\psi_{n}$ for $n$ gluons, according to (12), has
all these branch points and no other ones means that the energy
thresholds in the spectrum for $n$ gluons include all those for $n-1$
gluons (and maybe some new ones).
 This conclusion is in line with the
conjecture by G.Korchemsky [ 6 ], based on the Yang-Baxter equation
formalism applied to the gluon chain Hamiltonian in the limit
$N_{c}\rightarrow\infty$. Note, however, that our result is much
stronger: it applies to finite $N_{c}$ and the full gluon Hamiltonian
(8).

To conclude this section we present the explicit form of the kernel $W$
(for $i=3$)
\beq
(1/g)W(q_{1},q_{2},q_{3};q'_{1},q'_{3})=
\frac{q_{12}^{2}}{{q'_{1}}^{2}(q_{3}-q'_{3})^{2}}+
\frac{q_{13}^{2}}{{q'_{3}}^{2}(q_{2}-q'_{1})^{2}}-
\frac{q_{123}^{2}}{{q'_{1}}^{2}{q'_{3}}^{2}}-
\frac{q_{1}^{2}}{(q'_{1}-q_{2})^{2}(q_{3}-q'_{3})^{2}}
\eeq
where $q_{12}=q_{1}+q_{2}$ etc. This kernel (multiplied by
${q'_{1}}^{2}{q'_{3}}^{2}$ and with the opposite sign) was obtained by
J.Bartels as a vertex for transition from two to three reggeized gluons
in the four-gluon system, by studying the three-particle amplitude in the
triple Regge kinematics [ 7 ]. Our result may be considered as a
generalization to an arbitrary number of gluons in an arbitrary colour
state.

The described bootsrap of gluons  takes place for any number of colours.
However it leads to spectacular results in the limit
$N_{c}\rightarrow\infty$, which will be studied presently.

\section{The colour structure for $N_{c}\rightarrow\infty$. Multipomeron
solutions}

It is well-known that in the limit $N_{c}\rightarrow\infty$ the colour
group $SU(N_{c})$ does not differ from $U(N_{c})$ and the gluon may be
represented by a $q\bar q$ pair. Then the colour trace can be taken
by following quark lines. The leading contribution is obtained from
diagrams in which gluon lines do not cross. Taking as a target a $q\bar
q$ loop with external colourless sources (real or virtual photons) one
obtains a cylinder built on this loop as a dominant configuration, each
gluon iteracting only with his two neighbours. In this cylinder each
pair of neighbour gluons are in the adjoint representation. Indeed
crossing two neighbouring gluon lines, each represented by a $q\bar q$
pair, one encounters the configuration
\[\sum_{\gamma=1}^{N_{c}}q_{\alpha}^{(1)}{\bar
q}_{\gamma}^{(1)}q_{\gamma}^{(2)}{\bar q}_{\beta}^{(2)},\
\alpha,\beta=1,...N_{c}\]
which evidently transforms as a vector (it is a superposition of
antisymmetric and symmetric representations with equal weights). This
means that for the resulting chain Hamiltonian, each gluon interacting
only with his two neighbours, all $T_{i}T_{k}=-N_{c}/2$. As a result one
finds a colour independent chain Hamiltonian studied in [ 3-6 ].

Of course, the described colour structure, each pair of neighbours being
in the adjoint representation, is not the only one possible at
$N_{c}\rightarrow\infty$. The simplest configuration different from it
is the one in which pairs of gluons form a colourless state. For the
total colour zero it requires that the number of gluons $n$ be even.
Then the colour wave function takes the form
\beq
\chi_{a_{1},...a_{n}}=(N_{c}^{2}-1)^{-n/4}\delta_{a_{1}a_{2}}
\delta_{a_{3}a_{4}}...\delta_{a_{n-1}a_{n}}
\eeq
each $a_{i}$ taking the values from  1 to $N_{c}^{2}-1$.
It is important that in such a configuration gluons belonging to
different colourless pairs cannot couple to form a colourless pair or a
vector. Indeed the probability to find, say, the gluons 1 and 3 in a
colour
representation $R$ is determined by the projection $P^{(13)}_{R}\chi$
onto this representation. For the colourless state $R=(1)$ the projector
is
\beq
(P_{(1)}^{(13)})_{a_{1}a_{3},a'_{1}a'_{3}}=(N_{c}^{2}-1)^{-1}
\delta_{a_{1}a_{3}}\delta_{a'_{1}a'_{3}}
\eeq
Applying it to (19) we find
\beq
P_{(1)}^{(13)}\chi=
(N_{c}^{2}-1)^{-n/4-1}\delta_{a_{1}a_{3}}
\delta_{a_{2}a_{4}}...\delta_{a_{n-1}a_{n}}
\eeq
with the norm
\beq
\|P_{(1)}^{(12)}\chi\|^{2}=1/(N_{c}^{2}-1)^{2}
\eeq
Thus in the limit $N_{c}\rightarrow\infty$ $P_{(1)}^{(13)}\chi=0$, i.e.
the gluons 1 and 2 are never in a colourless state. Likewise for the
antisymmetric vector state
\beq
(P_{(N_{c}^{2}-1)_{A}}^{(13)})_{a_{1}a_{3},a'_{1}a'_{3}}=
c_{A}f_{a_{1}a_{3}a}f_{a'_{1}a'_{3}a}
\eeq
where $f$'s are the structure constants of $SU(N_{c})$ and $c_{A}$ is
the appropriate normalization constant. We find
\beq
P_{(N_{c}^{2}-1)_{A}}^{(13)}\chi=c_{A}f_{a_{1}a_{3}a}f_{a'_{2}a'_{4}a}
\delta_{a_{5}a_{6}}...\delta_{a_{n-1}a_{n}}
\eeq
and
\beq
\|P_{(N_{c}^{2}-1)_{A}}^{(13)}\chi\|^{2}=1/(N_{c}^{2}-1)
\eeq
The same result holds for the symmetric vector representation
$(N_{c}^{2}-1)_{S}$ with $f_{abc}\rightarrow d_{abc}$. Thus although the
probability to find the gluons 1 and 3 in a vector state is somewhat
higher than for a colourless state, it still goes to zero as
$N_{c}\rightarrow\infty$. In conclusion, in this limit
\beq
P_{(1)}^{(13)}\chi=
P_{(N_{c}^{2}-1)_{A}}^{(13)}\chi=
P_{(N_{c}^{2}-1)_{S}}^{(13)}\chi=0
\eeq

Now recall that for the total colour of $n$ gluons equal to zero the
Hamiltonian (8) can be represented in the form [ 3 ]
\beq
H=-(1/2)\sum_{i<k}^{n}(T_{i}T_{k})H_{ik}
\eeq
where
\beq
H_{ik}=t(q_{i})+t(q_{k})-2V_{ik}
\eeq
is a pair Hamiltonian for gluons $i$ and $k$ acting only on space
variables and infrared stable. For this pair of gluons in a given
colour representation $R$ we have
\beq
T_{i}T_{k}=(1/2)C_{2}(R)-N_{c}
\eeq
where $C_{2}$ is the quadratic Kasimir operator. Two gluons can combine
to form the following colour states: a colourless one $(1)$,
antisymmetric and symmetric vector states $(N_{c}^{2}-1)_{A,S}$, two
other symmetric representations, which we denote generically as $(S)$,
and two other antisymmetric representations $(A)$. The
values of the operator $T_{i}T_{k}$, found from (29) for representations
$(1),(N^{2}-1)_{A},(N^{2}-1)_{S},(S),(A)$ are respectively
\beq
-N_{c},\ -N_{c}/2,\ -N_{c}/2,\ \pm 1,\ 0
\eeq
{}From this we conclude that in the limit $N_{c}\rightarrow\infty$ two
gluons interact only when they are either in a colourless state or in a
vector one. Other symmetric states give a contribution of a relative
order $1/N_{c}$ and other antisymmetric states decouple altogether.

Returning to our colorless pair configuration (19) we observe that
gluons belonging to different pairs do not interact, since they cannot
be in a colourless or vector state. Therefore the Hamiltonian for such a
configuration reduces to a sum  of $n/2$
independent BFKL  pair Hamiltonians  (with $T_{2i-1}T_{2i}=-N_{c}$)
\beq
H=(N_{c}/2)\sum_{i=1}^{n/2}H_{2i-1,2i}
\eeq
The corresponding general solution of the homogeneous Schroedinger
equation is a product of $n/2$ BFKL pomerons
\beq
\psi_{n}(q_{1},...q_{n})=\prod_{i=1}^{n/2}\psi_{BFKL}(q_{2i-1},q_{2i})
\eeq
 with the energy which is the sum of the pomeron energies. The ground
state is evidently the BFKL pomeron one multiplied by $n/2$:
\beq
E_{n}=(n/2)E_{0},\ \ E_{0}=-(g^{2}N_{c}/\pi^{2})\ln 2
\eeq
Physically this solution is nothing but a normal $n/2$ cut. Our result
is thus the cut is an exact solution of the Hamiltonian for $n$
reggeized gluons in the large $N_{c}$ limit.

In the following we shall argue that the $n/2$ pomeron cut represents
the leading singularity associated with $n$ reggeized gluons in the
limit $N_{c}\rightarrow\infty$. Its simple structure and a simple form
for its interaction with a $q\bar q$ pair make it possible to sum
terms coming from arbitrary number of gluons and thus find the
leading contribution, which satisfies the unitarity in the limit of
large $N_{c}$. This will be done later, in Section 6., after we study
the chain Hamiltonian contribution with neighbouring gluons in the
vector colour state and show that this contribution is trivial.

Meanwhile we point out that the colourless pair configuration is not the
only alternative to the pure vector configuration coresponding to a
cylinder diagram. Evidently one may put the gluons in a mixed colour
state. Some of them may couple into colourless pairs:
 $(12),(34),...(2k-1,2k)$, the rest $n-2k$ gluons joining in a number of
cylinder
configurations.  The Hamiltonian for such a state splits into a sum of
$k$ independent BFKL Hamiltonians for colourless pairs and  chain
Hamiltonians for the rest gluons
\beq
H=(N_{c}/2)\sum_{i=1}^{k}H_{2i-1,2i}+(N_{c}/4)\sum_{i=2k+1}^{n}H_{i,i+1}
\eeq
with last and first coordinates identical for each cylinder. Solutions
to the corresponding
Schroedinger equation will be given by a product of $k$ BFKL pomerons
and solutions to the chain Hamiltonian problem for the rest gluons.

According to [ 10 ], for a configuration of $n$ gluons which splits into
$k$ colourless subconfigurations the colour factor is $N_{c}^{n+2-2k}$.
For the dominant configuration, the $n$-gluon cylinder ($k=1$) the
factor is $N_{c}^{n}$  and for the colourless pair configuration
($k=n/2$) it is $N_{c}^{2}$.
Thus it may look as if the latter configurations can safely be neglected
as $N_{c}\rightarrow\infty$. However, as we shall presently see, this is
not true.

\section{Solution to the gluon chain equation}

The peculiarity of the gluon chain in the limit $N_{c}\rightarrow\infty$
is that all gluons are in the vector colour state with respect to their
neighbours and this property is conserved when one bootstraps a pair of
them into a single gluon. This allows to apply the bootstrap mechanism
$(n-2)$ times consecutively to fuse all $n$ gluons into a pair in the
colourless state, which is a BFKL pomeron.

For the gluon chain all $T_{i}T_{k}=-N_{c}/2$ so that we need not bother
about the colour variables. Since the gluons are interacting only with
their neighbours, we have to retain only one of the two terms in (10)
(except for $n=3$). The chain Hamiltonian acting only on the space
variable is
\beq
H=\sum_{i=1}^{n}(t(q_{i})-V_{i,i+1}),\ \ n+1\equiv 1
\eeq
The inhomogeneous term for bootstrapping the gluons 1 and 2 can now be
directly written in terms of the space kernel $W$:
\[
F_{n}(q_{1},...q_{n})=-\int (d^{2}q'_{3}/(2\pi)^{3})
W(q_{2},q_{1},q_{3};q'_{1},q'_{3})\psi_{n-1}(q'_{1},q'_{3},q_{4},...q_{n})
\]\beq
-\int (d^{2}q'_{n}/(2\pi)^{3})
W(q_{1},q_{2},q_{n};q'_{1},q'_{n})\psi_{n-1}(q'_{1},q_{3},q_{4},...q'_{n})
+F_{n-1}(q_{1}+q_{2},q_{3},...q_{n})
\eeq
With this form of $F_{n}$ one finds that the solution to the
Schroedinger equation with the Hamiltonian (35) for $n$ gluons is given
by
\beq
\psi_{n}(q_{1},...q_{n})=g\psi_{n-1}(q_{1}+q_{2},q_{3},...q_{n})
\eeq
where $\psi_{n-1}$ is the solution of the same equation for $n-1$ gluon
and the inhomogeneous term $F_{n-1}$. The demonstration repeats that of
Section 2. except that now the $W$ terms substitute only two interaction
terms $V_{1n}$ and $V_{23}$ present in (35) for $n$ gluons by two terms
$V_{12,3}$ and $V_{12,n}$ which should appear for $n-1$ gluons.

We now repeat this procedure for
$\psi_{n-1}(q_{1}+q_{2},q_{3},...q_{n})$ bootstrapping another pair of
neighbours among the $n-1$ remaining gluons. Namely we write a
particular solution $\psi_{n-1}$ expressed via an arbitrary solution for
$n-2$ gluons
\beq
\psi_{n-1}(q_{1},...q_{n-1})=
g\psi_{n-2}(q_{1},...q_{i}+q_{i+1},...q_{n-1})
\eeq
Note that $i$ may be quite arbitrary. In particular we may choose $i=1$
or $i=n-1$ bootstrapping the gluon which is itself a result of fusion.
Continuing this process we finally express the solution to the original
equation for $n$ gluons through the solution $\psi_{2}$ for two gluons
in a colourless state (the BFKL pomeron):
\beq
\psi_{n}(q_{1},...q_{n})=g^{n-2}\psi_{2}(\tilde{q}_{1},\tilde{q}_{2})
\eeq
The momenta of the two gluons $\tilde{q}_{1},\tilde{q}_{2}$ are sums of
momenta of neighbouring gluons whose number and particular choice
depend on the bootstrapping process. The function $\psi_{2}
(\tilde{q}_{1},\tilde{q}_{2})$ satisfies the standard BFKL equation
with a certain inhomogeneous term $F_{2}(\tilde{q}_{1},\tilde{q}_{2})$.
It should evidently depend on two sums of momenta of $n$ initial gluons.
It is this term which ultimately selects the solution for the whole
tower of functions $\psi_{n},\ \psi_{n-1},\ ...\psi_{2}$

For the original equation for $n$ gluons all contributions which arise
as we successively represent $\psi_{n-1},\ \psi_{n-2},...$ appearing in
$F_{n},\ F_{n-1},....$ via $\psi_{2}$ represent a specific and rather
complicated inhomogeneous term. Thus what we have obtained is only a set
of particular solutions of the inhomogeneous Schroedinger equation.
We want to argue, however, that this set includes all the physically
interesting solutions. Any found solution, as mentioned, can be
characterized by the final inhomogeneous term
\beq
g^{n-2}F_{2}(\tilde{q}_{1},\tilde{q}_{2})\eeq where $\tilde{q}_{1}$
and $\tilde{q}_{2}$ are sums of momenta of neighbouring gluons in the
original $n$-gluon chain. This function represents the contribution to
$\psi_{n}$ in the lowest order in $g$, i.e. the coupling vertex for $n$
gluons to the target (projectile). Thus it is clear that should this
vertex have a more complicated structure (e.g. depend on three or more
momenta) then our set of solutions is too small. On the other hand, for
any  vertex of the form (40) we can consecutively construct $\psi_{2},
\psi_{3},....\psi_{n}$ using the bootstrapping procedure for the gluons
whose momenta are summed in the arguments of the function (40).

It is difficult to say something definite about the coupling of $n$
gluons to a general hadronic target because of the nonperturbative
effects. However in the case when one can study the coupling
perturbatively, namely, for the photon target, the coupling vertex,
as will be shown in the next section, indeed depends only on two momenta
which are sums of the gluon momenta, i.e. has the form (40). Thus at
least in this case the found set of solutions seems to cover the
physical ones. As to other contributions appearing in the inhomogeneous
term $F_{n}$ in the course of bootstrapping, they all take into account
processes when first two gluons couple to the target which then
consecutively split into more and more gluons until their number becomes
$n$ and they start to interact pairwise with the BFKL kernel. This
picture was observed by J.Bartels in the four-gluon system [ 7 ],
upon the study of appropriate cuts of the three-particle amplitude in
the triple Regge-kinematics.

Thus, if our argument is correct, the $n$-gluon chain in the overall
colorless state is equivalent to a single BFKL pomeron.
Its relative weight is determined by the factor $(g^{2}N_{c})^{n}$
which comes from the coupling to the external source and colour factor.
The BFKL pomeron itself corresponds to $n=2$ and his weight is
$g^{4}N_{c}^{2}$. Evidently all other gluon chains with $n\geq 3$ give
corretions to the BFKL pomeron coupling of the relative order
$(g^{2}N_{c})^{n}$. The parameter $g^{2}N_{c}$ is however assumed small
in this essentially perturbative approach. As a result,
in the limit $N_{c}\rightarrow\infty$, $g^{2}N_{c}\rightarrow 0$ we
can completely neglect all cylinder configurations and all
solutions to the $n$-gluon
problem reduce to $n/2$ non-interacting BFKL pomerons,
which is nothing but the normal $n/2$ cut contribution. Since no
interaction remains between pomerons for large $N_{c}$, it is not
difficult to sum this leading contribution for all $n$ and obtain a
unitary amplitude. To do that we have to study the coupling of $n$
gluons to the projectile (target). This will be done in the next
section.

\section{The $n$-gluon-photon coupling}

As mentioned, we choose  photons (real or virtual) as the target and
projectile , since in this case we can study the coupling to gluons
perturbatively. To separate the coupling we consider the
$\gamma^{\ast}\gamma$ forward scattering amplitude corresponding to
$n$-gluon exchange. The procedure we adopt closely follows the
well-known one for the two-gluon exchange (see [ 11 ]), so that we shall
be brief.

Let the momentum of the virtual photon (projectile) be $q$ and that of
the real photon (target) be $p$. We choose a system where
$p_{+}=q_{-},\ p_{-}=p_{\bot}=q_{\bot}=0$. Then $\nu=pq=p_{+}q_{-}=
p_{+}^{2}\rightarrow\infty$, $x=-q^{2}/(2\nu)$ and $q_{+}=-xp_{+}$. We
shall consider the case of $x<<1$, so that in what follows we always
neglect $x$ as compared to quantities of the order unity. The forward
scattering amplitude corresponding to the $n$-gluon exchange is written
in the form
\beq
iA(p,q)=(1/n!)\int\prod_{k=1}^{n-1}(d^{4}q_{k}/(2\pi)^{4})i\Gamma_{p}
(q,q_{i})i\Gamma_{t}(p,q_{i})\prod_{k=1}^{n}(i/q_{k}^{2})
\eeq
where $\sum_{k=1}^{n}q_{k}=0$. The $\Gamma$'s are 4-dimensional vertices
for the interaction of the target ($t$) and projectile ($p$) with $n$
gluons. Vector multiplication in colour and space gluon indeces is
understood. Vector indeces should also be associated with the
interacting photons. As shown in [ 11 ] at high energies one
can factorize the vector summation in space indeces conserving only the
longitudinal components of the exchanged gluons, which amounts to
changing each metric tensor according to
\beq
g_{\alpha\beta}\rightarrow p_{\alpha}q'_{\beta}/\nu
\eeq
where $q'=q+xp$ and $\alpha(\beta)$ refers to the projectile (target).
As to the photon vector indeces, we project onto its longitudinal and
transversal components by means of the standard projectors
\beq
P_{L}^{\alpha\beta}=(q^{2}/\nu)p^{\alpha}p^{\beta};\ P_{\bot}
^{\alpha\beta}=(1/2)g_{\bot}^{\alpha\beta}
\eeq

We are interested in a kinematical situation when the gluons are
reggeized. It means that all intermediate states in the two $\Gamma$'s
should have finite masses (not growing with $\nu$). From that we
conclude that $q_{i\pm}$ are small and $q_{i}^{2}\simeq q_{i\bot}^{2}$.
Also one concludes that the projectile vertex $\Gamma_{p}$ does not
depend on $q_{i-}$ and the target one $\Gamma_{t}$ does not depend on
$q_{i+}$. This makes it possible to present (41) as a multiple two
dimensional integral with a factorized integrand
\beq
A(p,q)=(4\nu/n!)i^{n+1}\int\prod_{k=1}^{n-1}(d^{2}q_{k}/(2\pi)^{2})F_{p}
(q,q_{i})F_{t}(p,q_{i})\prod_{k=1}^{n}(1/q_{k\bot}^{2})
\eeq
Applying (42) we find for the projectile part
\beq
F_{p}(q,q_{i})=(1/2q_{-})(q_{-}/\nu)^{n}\int
\prod_{k=1}^{n-1}(dq_{k+}/(2\pi))\Gamma_{p}^{\alpha_{1}...\alpha_{n}}
(q,q_{i})p_{\alpha_{1}}...p_{\alpha_{n}}
\eeq
where $q_{i-}=0$. This formula defines the 2-dimensional vertex for the
interaction of the projectile with $n$ gluons. For the target we obtain
a similar formula with $p\rightarrow q'$. It is these vertices that
enter the $n$ gluon equation as inhomogeneous terms.

To calculate the integrals over $q_{i+}$ in (45) we change to variables
\[ s_{k}=(q+\sum_{i=1}^{k}q_{i})^{2},\ \ k=1,...n-1\]
and deform the Feynman integration contour in each $s_{k}$ to close
around the cut along the positive axis
\beq
F_{p}(q,q_{i})=(1/2)\nu^{-n}\int\prod_{k=1}^{n-1}(ds_{k}/4\pi){\rm disc}
\Gamma_{p}(q,s_{i})
\eeq
The multiple discontinuity in all variables $s_{1},...s_{n-1}$ should be
taken. It can be calculated by using the unitarity condition and
inserting some intermediate states. In the lowest order in the coupling
constant one retains only the simplest $q\bar q$ state in each
discontinuity.  The transition from the $i-1$th to the $i$th
intermediate state is accomponied by the emission of the $i$th gluon,
$i=2,...n-1$, the gluon $1(n)$ present in the initial (final)
state. Each gluon may be emitted either by the quark or by the
antiquark. This gives $2^{n}$ different contributions.

This result exactly corresponds to calculating the $q\bar q$ loop with
$n$ gluons attached to it in a specific manner. All $q$ and $\bar q$
denominators $(m^{2}-k^{2})^{-1}$ should be changed to $2\pi i\delta
(m^{2}-k^{2})$ except the ones which connect the photon vertices with
gluons 1 and $n$. Also the gluons should follow in the rising order both
on the $q$ and $\bar q$ lines. If the numbers of the gluons emitted from
the $q$ line are $i_{1},...i_{n_{1}}$ we should have
\beq
i_{1}<i_{2}<...<i_{n_{1}}
\eeq
and similarly for the rest $n-n_{1}$ gluons attached to the $\bar q$
line. With the order of gluons emitted from $q$ and $\bar q$ lines
fixed, different contributions are obtained only when some of the gluons
change their source from $q$ to $\bar q$ or vice versa. With $n_{1}$
gluons emitted from $q$ and $n_{2}=n-n_{1}$ gluons emitted from $\bar
q$, the total number of different contributions is $C_{n}^{n_{1}}$.
Summed over $n_{1}$ it gives $2^{n}$ as it should be.

With all but two gluon propagators changed to $\delta$-functions the
integrations over the initial quark momentum $k_{0+}$  and $q_{i+}\
i=1,...n-1$ become trivial. To simplify the presentation we start with
the case of an Abelian gauge group, i. e. when there are no colour
indeces. Then the only preliminary step left is to take the spinor trace
in the loop. The easiest case is  the longitudinal photon. Then all
vertices in the loop are changed to $\hat{p}=p_{+}\gamma_{-}$.
Correspondingly of every propagator numerator only the part
$k_{i-}\gamma_{+}$ is left. All "-" components of the quark or antiquark
momenta are equal because $q_{i-}<<k_{i-}$. The longitudinal trace thus
becomes
\beq
T_{L}=-(q^{2}/\nu^{2})2^{n+3}p_{+}^{n+2}k_{0-}^{n_{1}+1}(k_{0-}-q_{-}
)^{n_{2}+1}
\eeq
where $n_{1}(n_{2})$ is the number of gluons emitted from $q({\bar q})$
and $k_{0}$ is the quark's momentum before all emissions. For the
transverse photon we find two transverse photon vertices $\gamma_{\bot}$
which require that in  the adjoining propagators the transverse
part $\hat{k}_{i\bot}$ or the mass term should be retained.
The transverse trace results
\beq T_{\bot}=
-2^{n+1}p_{+}^{n}k_{0-}^{n_{1}-1}(k_{0-}-q_{-}
)^{n_{2}-1}(m^{2}+(k_{0-}^{2}+(q_{-}-k_{0-})^{2})(k_{0}\tilde{k})_{\bot})
\eeq
where $\tilde{k}$ is the quark's momentum after all emissions.

Let us take the configuration when the gluons $1,...n_{1}$ are emitted
from the quark and the gluons $n_{1}+1,...n$ are emitted from the
antiquark. Performing the integrations over $k_{0+}$ and $q_{i+},\
i=1,...n-1$ we find the contribution to the longitudinal vertex $F_{L}$
\[
F_{L}^{(n_{1})}(q_{i\bot})=4i(-1)^{n_{1}}q^{2}e^{2}g^{n}
\sum_{f=1}^{N_{f}}Z_{f}^{2}\int_{0}^{1}d\alpha(\alpha (1-\alpha))^{2}\]
\beq
\int (d^{2}k/(2\pi)^{3})
(k_{\bot}^{2}+\epsilon_{f}^{2})^{-1}((k+\sum_{i=1}^{n_{1}}q_{i})_{\bot}^{2}
+\epsilon_{f}^{2})^{-1}
\eeq
where
\beq
\epsilon_{f}^{2}=m_{f}^{2}-\alpha (1-\alpha)q^{2}
\eeq
The summation goes over quarks of different flavours $f$ with masses
$m_{f}$ and charges $Z_{f}e$;
$\alpha$ is the scaling variable for the quark: $k_{0-}=\alpha
q_{-}$. The transverse vertex has a similar structure
\[
F_{\bot}^{(n_{1})}(q_{i\bot})=i(-1)^{n_{1}}e^{2}g^{n}
\sum_{f=1}^{N_{f}}Z_{f}^{2}\int_{0}^{1}d\alpha
\int (d^{2}k/(2\pi)^{3})\]\beq
(m_{f}^{2}+(\alpha^{2}+
(1-\alpha)^{2})
(k,k+\sum_{i=1}^{n_{1}}q_{i})_{\bot})
(k_{\bot}^{2}+\epsilon_{f}^{2})^{-1}((k+\sum_{i=1}^{n_{1}}q_{i})_{\bot}^{2}
+\epsilon_{f}^{2})^{-1}
\eeq
Both contributions depend only on the sum $\sum_{i=1}^{n_{1}}q_{i}$ of
the gluon momenta  emitted from the quark (or antiquark). The total
vertices are obtained upon summing over all different
distributions of gluons between the quark and antiquark.

Now we introduce colour variables. We first consider the case when
gluons form colourless pairs, so that their colour wave function
has the structure (19). In this case the colour matrices of the quark
loop $t^{a},\ a=1,....N_{c}$ should be summed pairwise over $a$.
Diagrammatically it is equivalent to joining the corresponding loop
vertices by a gluon line (only in colour space). In the limit
$N_{c}\rightarrow\infty$ the leading contribution will come from
configurations in which all these $n/2$ gluon colour lines do not cross,
i.e the loop with these spurious colour lines is planar. In such a
planar loop all different pairs $t^{a}t^{a}$ can be subsequently put
together and summed over $a$ to give $(N_{c}^{2}-1)/N_{c}\sim N_{c}/2$
each. The colour trace then becomes
\beq
c_{n}=N_{c}(N_{c}/2)^{n/2}
\eeq
Turning to our derivation of the vertex without colours, we observe that
with the order of gluons fixed by the condition (47) , the gluon colour
lines joining the pairing loop vertices never cross. In fact, suppose
that two lines connecting quark vertices $i,i+1$ and $k,k+1$, $i<k$,
cross. This may only happen if $k+1<i+1$, which is impossible. Similarly
one shows that gluons lines do not cross if they start on the quark line
and finish on the antiquark line. As a result, introduction of colour
for the colourless pair configuration does not change the vertices found
for the Abelian case except for the overall colour factor $c_{n}$, Eq.
(53).

A different result is obtained for the configuration in which gluons are
in the adjoint representation with their neighbours and form a cylinder
with the two $q\bar q$ loops as bases. Each gluon together with the
quark from the target and antiquark from the projectile (or vice versa)
then contribute a factor $N_{c}$, giving an overall colour factor
$N_{c}^{n}$ (for the amplitude). Comparing with the Abelian case we find
that we have to retain only a part of the configurations which
correspond to a fixed order of the gluons along the $q\bar q$ loop (say,
$1,...n$) arbitrarily divided between the quark and antiquark. As a
result only $n$ terms survive for each $n_{1}\neq 0,n$ and not
$C_{n}^{n_{1}}$ as in the previous cases. Still the obtained vertex
always depends on only two momenta $\tilde{q}_{1}$ and $\tilde{q}_{2}$
which are sums of the momenta of neighbouring gluons (in fact, for the
forward scattering $\sum_{i=1}^{n}q_{i}=0$ so that
$\tilde{q}_{1}+\tilde{q}_{2}=0$ and the vertex actually depends on only
one momentum). This fact was used in the previous section to argue
that the solution obtained for the chain Hamiltonian was general enough.

\section{The photon structure function at low $x$}

With the leading contribution given by $n/2$ non-interacting pomerons
and the verteces for the $n$-gluon coupling to the target and projectile
known, we can proceed to sum  contributions from all $n$ to obtain a
unitary description for the $\gamma^{\ast}\gamma$ scattering. We assume
that gluons are paired into BFKL pomerons in the order
$(12),(34),...(n-1,n)$. In fact we have $(n-1)!!$ different orderings,
which all give the same contribution. This changes the symmetry factor
according to
\beq
(n-1)!!/n!=1/(2^{n/2}(n/2)!)
\eeq
The factor $(1/2)^{n/2}$ corresponds to the symmetry factor $1/2$ for
each pair; $1/(n/2)!$ is the correct symmetry factor for $n/2$ pomerons.

To perform the summation over $n$ we first present the sum of all
contributions (51) or (52) to the vertices coming from different
distributions of the gluons between $q$ and $\bar q$ in a convenient
manner. We use the approach proposed in [ 12 ] based on the impact
parameter space. One writes
\[
(k_{\bot}^{2}+\epsilon_{f}^{2})^{-1}=\int\,(d^{2}r/2\pi)\,
{\rm K}_{0}(\epsilon_{f} r)
\]
where ${\rm K}_{0}$ is the McDonald function. Then the contribution (51)
from the given gluon distribution to the longitudinal vertex is
rewritten as (with the colour factor (53))
\[
F_{L}^{(n_{1})}(q_{i\bot})=4ic_{n}(-1)^{n_{1}}q^{2}e^{2}g^{n}\]\beq
\sum_{f=1}^{N_{f}}Z_{f}^{2}
\int_{0}^{1}d\alpha(\alpha (1-\alpha))^{2}\int (d^{2}r/(2\pi)^{3})
 {\rm K}_{0}^{2}(\epsilon_{f} r)
\exp (ir\sum_{i=1}^{n_{1}}q_{i})
\eeq
Following [ 12 ] we introduce the longitudinal $q\bar q$ density of the
projectile in the $\alpha  ,r$ space
\beq
\rho_{L}(\alpha,r)=\sum_{f=1}^{N_{f}}Z_{f}^{2}(\alpha
(1-\alpha))^{2}{\rm K}_{0}^{2}(\epsilon_{f} r) \eeq
 The sum over all distributions
results as \[
F_{L}^{(n_{1})}(q_{i\bot})=4ic_{n}(-1)^{n_{1}}q^{2}e^{2}g^{n}\]\beq
\int_{0}^{1}d\alpha\int (d^{2}r/(2\pi)^{3})\rho_{L}(\alpha,r)
\prod_{i=1}^{n}(\exp (irq_{i})-1)
\eeq
In the same way for the transverse vertex we introduce the transverse
density of the projectile
\beq
\rho_{\bot}(\alpha,r)=
\sum_{f=1}^{N_{f}}Z_{f}^{2}(m_{f}^{2}{\rm K}_{0}^{2}(\epsilon_{f} r)
+(\alpha^{2}+(1-\alpha)^{2})\epsilon_{f}^{2}{\rm K}_{1}^{2}
(\epsilon_{f} r))
\eeq
The transverse vertex aquires the form
\[
F_{\bot}^{(n_{1})}(q_{i\bot})=ic_{n}(-1)^{n_{1}}e^{2}g^{n}\]\beq
\int_{0}^{1}d\alpha
\int (d^{2}r/(2\pi)^{3})\rho_{\bot}(\alpha,r)
\prod_{i=1}^{n}(\exp (irq_{i})-1)
\eeq
Analogous expressions are obtained for the target (real and transverse)
photon with $q^{2}=0$.

Now we can calculate the contribution to the amplitude coming from the
exchange of $n/2$ non-interacting pomerons. We have only to change pairs
of gluon propagators in (44) by BFKL pomerons taken in the energy
representation. Let $l=q_{1}+q_{2}=q'_{1}+q'_{2}$ be the total momentum
of the two gluons 1 and 2 forming a pomeron, $k=(1/2)(q_{1}-q_{2})$ and
$k'=(1/2)(q'_{1}-q'_{2})$ be their final and initial relative momenta.
Then we have to change
\beq
(2\pi)^{2}\delta^{3}(k-k')/q_{1}^{2}q_{2}^{2}\rightarrow f(\nu,l,k,k')
\eeq
where\beq
f(\nu,l,k,k')=\int\,(dE/2\pi
i)\nu^{-E}\psi_{2}(E,l,k,k')/q_{1}^{2}q_{2}^{2} \eeq
and $\psi_{2}(E,l,k,k')$ is the solution of Eq. (1) for two gluons in a
colourless state with the inhomogeneous term
$F_{2}=(2\pi)^{2}\delta^{2}(k-k')$. The integration in (61) runs along
the
imaginary axis to the left of all singularities of the integrand in $E$.
We find for the contribution from the $n/2$ pomeron exchange to the
$\gamma^{\ast}\gamma$ forward scattering amplitude (for the transverse
projectile photon) \[
A_{\bot}^{(n)}(p,q)=-4i\nu e^{4}N_{c}^{2}
(-g^{4}/16)^{n/2}(1/(n/2)!)
\int_{0}^{1}d\alpha \int_{0}^{1}d\beta
\]\[
\int\,d^{2}R \int d^{2}r \int d^{2}r'(2\pi)^{-6}
\int\prod_{i=1}^{n/2}(d^{2}ld^{2}kd^{2}k'(2
\pi)^{-6} f(\nu,l_{i},k_{i},k'_{i})\exp (il_{i}R))\]\beq
\rho_{\bot}(\alpha,r)\rho_{\bot}(\beta,r',\epsilon_{f}=m_{f})
\prod_{i=1}^{n}(\exp (irq_{i})-1) (\exp (-ir'q'_{i})-1)
\eeq
The $R$ integration takes into account the condition $\sum_{i=1}^{n/2}
l_{i}=0$

 The integration over the pomeron momenta gives
\[\int\,(d^{2}ld^{2}kd^{2}k'/(2\pi)^{6})
f(\nu,l,k,k')\exp(ilR) (\exp iq_{1}r -1)
 (\exp iq_{2}r -1)\]\beq(\exp (-iq'_{1}r') -1)
(\exp (-iq'_{2}r') -1)=
4(2\pi)^{-4}f(\nu,R+(1/2)(r-r'),r,r')
\eeq
where $f(\nu,R,r,r')$ is the BFKL correlator in the impact parameter
space found in [ 13 ]. Retaining only the $s$-wave contribution dominant
at large $\nu$
\beq
f(\nu,R,r,r')=\int
d\kappa\frac{\kappa^{2}\nu^{-E(\kappa)}}{(\kappa^{2}+1/4)^{2}}
\int d^{2}r_{0}
(\frac{r}{r_{10}r_{20}})^{1+2i\kappa}
(\frac{r'}{r'_{10}r'_{20}})^{1-2i\kappa}
\eeq
where \[r_{10}=R+r/2-r_{0},\ r_{20}=R-r/2-r_{0},\ r'_{10}=r'/2-r_{0},\
r'_{20}=-r'/2-r_{0}\]
and $E(\kappa)$ is the BFKL pomeron energy
\beq
E(\kappa)=(g^{2}N_{c}/2\pi^{2})({\rm Re}\,\psi(1/2+i\kappa)-\psi(1))
\eeq

 Under the sign of integrals over $\alpha,\,\beta\,R$, $r$ and $r'$ in
(62) we find the $n/2$th
power of $f(\nu,R,r,r')$. Evidently the contribution has an eikonal
form. Summing
over $n$ we finally obtain the total amplitude for the transverse
projectile as an integral of the known functions
\[
A_{\bot}(p,q)=4i\nu e^{4}N_{c}^{2}
\int_{0}^{1}d\alpha \int_{0}^{1}d\beta
\]\beq
\int\,d^{2}R \int d^{2}r \int d^{2}r'(2\pi)^{-6}
\rho_{\bot}(\alpha,r)\rho_{\bot}(\beta,r',q^{2}=0)
(1-\exp (-\frac{g^{4}}{4(2\pi)^{4}}f(\nu,R,r,r')))
\eeq
For the longitudinal projectile photon one has to change
$\rho_{\bot}(\alpha,r)$ to $\rho_{L}(\alpha,r)$.

The amplitude (66) represents the leading contribution to the
$\gamma^{\ast}\gamma$ forward scattering amplitude in the limit
$N_{c}\rightarrow\infty$. It has an eikonal structure and is evidently
unitary.

At large $\nu$  small values of $\kappa$ dominate in (64) where
\beq
E(\kappa)=E_{0}+a\kappa^{2},\ \  a=(7/2)\zeta (3)
\eeq
and $E_{0}$ is negative (see (33)). This gives the well-known large
factor $\nu^{|E_{0}|}$ in $f$. From the structure of $f$ it is then
clear that the dominant contribution comes from the region of small $r$
and $r'$. In this region  $f(\nu,R,r,r')$ results a function of only
two scalar variables $\nu$ and $z=R^{2}/rr'>>1$:
\beq
f(\nu,R,r,r')_{r,r'<<R}\simeq \frac{16\pi^{3/2}\nu^{|E_{0}|}}
{(a\ln \nu)^{3/2}}z^{-1}\ln z\exp (-\frac{\ln^{2} z}{a\ln\nu})
\eeq
where $a$ is given in (67).

Passing in (66) to the integration over $z$ we find that in the
asymptotical region $\nu\rightarrow\infty$ the amplitude completely
factorizes
\beq
A_{\bot}(p,q)=G_{\bot}(q^{2})F(\nu )G_{\bot}(p^{2})
\eeq
Here $F(\nu)$ is the contribution from all exchanged pomerons (the
"Froissaron") and $G_{\bot}(q^{2})$ and $G_{\bot}(p^{2})$ are its
 effective couplings
to the projectile and target respectively. For the projectile
\beq
G_{\bot}(q^{2})=(e^{2}N_{c}/4\pi)\int_{0}^{1}d\alpha\int_{0}^{\infty}
r^{2}dr\rho_{\bot}(\alpha,r)
\eeq
In the limit $Q^{2}=-q^{2}\rightarrow\infty$ we find
\beq
G_{\bot}(q^{2})=\frac{e^{2}N_{c}a_{1}}{2Q}\sum_{f}Z_{f}^{2}
\eeq
where $a_{n}$ are numbers defined by
\beq
a_{n}=\int_{0}^{\infty}r^{2}dr{\rm K}^{2}_{n}(r)
\eeq
For the real photon as a target ($p^{2}=0$)
\beq
G(0)_{\bot}=\frac{e^{2}N_{c}(a_{0}+2a_{1})}{4\pi}\sum_{f}Z_{f}^{2}/m_{f}
\eeq

{}From (71) we observe that the cross-section falls only as $1/Q$ at large
$Q^{2}$ instead of the standard scaling behaviour $1/Q^{2}$. This means
that the structure function rises roughly as $Q$.

The $\nu$ dependence is determined by the Froissaron $F(\nu)$:
\beq
F(\nu)=(i\nu/\pi)\int_{1}^{\infty}dz(1-\exp(-bz^{-1}\ln z\exp (-\xi
\ln^{2} z)))
\eeq
where according to (68)
\[ \xi=(a\ln\nu)^{-1},\ \ b=g^{4}\nu^{|E_{0}|}\xi^{3/2}/(4\pi^{3/2})\]
To crudely estimate the behaviour of $F(\nu)$ at large $\nu$ we
approximately determine the point $x_{0}>>1$ where the exponent becomes
small: $ x_{0}\sim \nu^{\lambda}$ where
\beq
\lambda=(1/2)(\sqrt{a^{2}+4a|E_{0}|}-a)
\eeq
Then neglecting the exponential function in the region $x<x_{0}$
we obtain
\beq
F(\nu)\simeq (i/\pi)\nu^{1+\lambda}
\eeq
Thus after summing over all pomeron exchanges
 the cross-section remains growing as a power $\lambda$ of the c.m.
energy squared (somewhat smaller than the original power $|E_{0}|$).
Correspondingly the small $x$ behaviour of the structure function is a
power growth
\beq F_{2}(Q^{2},x)\sim Q^{1+2\lambda}x^{-\lambda}\eeq

THe obtained undesirable features of the seemingly unitary
$\gamma^{\ast}\gamma$
amplitude are evidently related to the scaling invariant character of
the theory. With the gluon mass equal to zero and no confinement the
projectile and target can interact at arbitrary large distances. We
hope that introduction of the running coupling and the QCD scale
$\Lambda$ will lead to a more realistic picture.

\section {Conclusions}

We have generalized the bootstrap relation to the case of $n$ reggeized
gluons. It allows to conclude that the intercept generated by $n$ gluons
cannot be lower than for $n=2$. In particular the odderon intercept
($n=3$) cannot be lower than the pomeron one, so that the variational
estimate obtained in [ 14 ] is too low. With a $q\bar q$ loop as the
external source, the odderon possesses the same intercept as the BFKL
pomeron.

The most far-reaching consequences seem to follow from the bootstrap in
the limit $N_{c}\rightarrow\infty$ for the gluon chain with only
 neighbours interacting. We have given arguments that all physically
interesting solutions of this system reduce to a single BFKL pomeron.
This conclusion means that the BFKL pomeron coincides with the
topological pomeron, which corresponds to a sum of all diagrams with a
topology of a cylinder in the large $N_{c}$ limit [ 15 ].

The leading  contribution from the $n$-gluon system in the
large $N_{c}$ limit comes from $n/2$ non-interacting BFKL pomerons. We
have summed this contribution to the $\gamma^{\ast}\gamma$ forward
scattering amplitude for all $n$ and obtained a closed formula of an
eikonal form expressed through the known functions. Crude estimates
reveal that the resulting cross-section continues to grow as a power with
energy and falls as $1/Q$ with the virtuality  of the photon projectile
 $q^{2}=-Q^{2}$, which leads to a gross violation of scaling.

These results were obtained in the standard framework of the BFKL theory,
namely, for an infrared regularized theory with a small fixed coupling
(and hence no confinement). One should be most cautious in trying to
extend them to the realistic QCD. For such an extension the running
coupling constant and  the corresponding QCD scale have to be introduced.

\section{Acknowledgements}

The author expresses his deep gratitude to Prof. L.N.Lipatov for most
illuminating discussions, to Prof. V.D.Liakhovski for
consultations on the properties of the group $SU(N)$ and to Prof.
I.V.Komarov who drew his attention to the paper [ 6 ].

\newpage
\section{References}

1. J.Bartels, Nucl. Phys. {\bf B175} (1980) 365.\\
2. J.Kwiecinski and M.Praszalowicz, Phys. Lett. {\bf B94} (1980) 413.\\
3. L.N.Lipatov, Phys. Lett. {\bf B309} (1993) 394.\\
4. L.N.Lipatov, JETP Letters {\bf 59} (1994) 571.\\
5. L.D.Faddeev and G.P.Korchemsky, Stony Brook preprint ITB-SB-94-14
(1994).\\
6. G.P.Korchemsky, preprint HEP-PH/9501232.\\
7. J.Bartels, Z.Phys. {\bf C60} (1993) 471.\\
8. V.S.Fadin, E.A.Kuraev and L.N.Lipatov, Phys. Lett. {\bf B60} (1975)
50.\\
I.I.Balitsky and L.N.Lipatov, Sov.J.Nucl.Phys. {\bf 15} (1978) 438.\\
9. L.N.Lipatov, Yad. Fiz. {\bf 23} (1976) 642.\\
10. G.t'Hooft, Nucl. Phys. {\bf B72} (1974) 461;\\
11. L.N.Lipatov and G.V.Frolov, Yad. Fiz. {\bf 13} (1971) 588.\\
12. N.N.Nikolaev and B.G.Zakharov, Z.Phys. {\bf C49} (1991) 607.\\
13. L.N.Lipatov, Sov. Phys. JETP {\bf 63} (1986) 904.\\
14. P.Gauron, L.N.Lipatov and B.Nicolescu, Z.Phys. {\bf C63} (1994)
253.\\
15. G.Veneziano, Phys. Lett. {\bf B52} (1974) 220; Nucl. Phys. {\bf
B117} (1976) 519.\\

 \end{document}